\documentclass{jabad}
\newcommand{\beq}{\begin{equation}}
\newcommand{\eeq}{\end{equation}}
\newcommand{\barr}{\begin{eqnarray}}
\newcommand{\earr}{\end{eqnarray}}

\begin{document}


\title{Noncommutativity as a mapping of paths}

 \newcommand{\shortauthor}{{J.M. Carmona, J.L. Cort\'es, J. Indur\'ain
 and D. Maz\'on}}

 \author{Jos\'e Manuel Carmona\footnote{jcarmona@unizar.es},
 Jos\'e Luis Cort\'es\footnote{cortes@unizar.es}, Javier
 Indur\'ain\footnote{indurain@unizar.es}  and Diego Maz\'on\footnote{dmazon@unizar.es
}}
\begin{center}
  {Departamento de F\'{\i}sica Te\'{o}rica, Facultad de Ciencias,\\
  Universidad de Zaragoza, 50009 Zaragoza, Spain}
\end{center}

\begin{abstract}
A reinterpretation of noncommutativity as a mapping of paths is
proposed at the level of quantum mechanics.

\end{abstract}

\KeysAndCodes{Noncommutativity, path integral}{11.10.Nx;
03.65.-w; 02.40.Gh}


\section{Introduction}

In memory of Julio Abad, with whom one of
the authors (J.L.C.) shared more than thirty years of professional
experience, another one (J.M.C.) fifteen years and the other two the
beginning of their research career. During all these years we have
been lucky enough to experience Julio's kindness, his availability to
speak with him at any time, his joy for physics and love for good books. We are
proud to contribute to a volume in his honor with an article that, we
like to think, he might well have enjoyed.

Noncommutative geometry was considered and developed as a mathematical
generalization of commutative geometry, with an application to
physics, during the 1980s, mainly from Alain Connes approach to gauge
theories~\cite{connes}.

Noncommutativity had entered physics, however, much earlier, with the
advent of quantum mechanics. In ordinary quantum mechanics,
position and momentum are described by noncommutative self-adjoint
operators, but the geometry of space is the usual
one. There are however arguments suggesting that in a
quantum theory including gravity, position measurements will be
problematic at the Planck length and the geometry of space will have
to be changed at these small scales~\cite{lizzi}.

The arising of noncommutative spaces in string theory~\cite{string}
has in fact given support to this idea and led to a stronger interest
in the study of physical systems on a noncommutative geometry, in
particular of the quantum mechanics of particles on such
spaces~\cite{NCQM}.
More recently, it has been shown that the effective low-energy limit of 3-d
quantum gravity coupled to quantum matter (non gravitational) fields
(i. e., when the gravitational degrees of freedom are integrated out)
is equivalent to a quantum field theory on a 3-d noncommutative
spacetime ~\cite{freidel}.

In this paper we treat a very simple problem, the harmonic
oscillator on the noncommutative plane, and try to give a new
perspective of its analogies and differences with respect to the
standard quantum harmonic oscillator on the commutative plane, by
means of a path integral approach to both systems.


\section{Noncommutative Quantum Mechanics}\label{sec2}

The spectrum of a harmonic oscillator on the noncommutative plane was
analyzed in Refs.~\cite{NCQMOsc}. Let us review the main results. The
Hamiltonian of the system is
\beq
H=\frac{1}{2m}(\tilde p_1^2+\tilde p_2^2)+\frac{1}{2}m\omega^2 (\tilde
q_1^2+\tilde q_2^2),
\eeq
with $[\tilde q_i,\tilde p_j]=i\hbar \delta_{ij}$, $[\tilde q_1,\tilde
  q_2]=i \tilde \theta$, where $\tilde \theta$ is the noncommutativity
  parameter. It is convenient to rescale phase space coordinates so
  that the Hamiltonian is written in terms of dimensionless variables:
\beq
q_i=\sqrt{\frac{m\omega}{\hbar}}\tilde q_i\, , \quad
p_i=\frac{1}{\sqrt{m\omega\hbar}}\tilde p_i\, ,
\eeq
so that $[q_i,p_j]=i\delta_{ij}$, $[q_1,q_2]=i\theta$, where
\beq
\theta=\frac{m\omega}{\hbar} \tilde{\theta}
\eeq
is the dimensionless rescaled noncommutativity parameter, and the
Hamiltonian is re-expressed as
\beq
H=\frac{\hbar\omega}{2}(p_1^2+p_2^2+q_1^2+q_2^2).
\eeq

In the following we will omit the $\hbar$ factors. A simple way to
solve this system is to make a Darboux transformation $(q_i, p_i) \mapsto (Q_i, P_i)$ such that the modes of vibration are decoupled. Defining the quantities
\beq
\lambda_\pm = \sqrt{1+\frac{\theta^2}4}\pm \frac \theta 2 = (\lambda_\mp)^{-1}\,,
\eeq
then the change of variables is given (up to rotations in the $\{Q_1, P_1\}$ and $\{Q_2,P_2\}$
planes) by
\beq
\xi = \left(
\begin{array}{c}
 q_1 \\ q_2 \\ p_1 \\p_2
\end{array}
\right)
\, = \,
\sqrt{\frac{\lambda_+}{1+\lambda_+^2}}\left(
\begin{array}{cccc}
 \lambda_+ & \lambda_- & 0 & 0\\
0 & 0 & \lambda_+ & -\lambda_-\\
0 & 0 & 1 & 1 \\
-1 & 1 & 0 & 0
\end{array}
\right)\left(
\begin{array}{c}
 Q_1 \\ Q_2 \\ P_1 \\ P_2
\end{array}
\right)
\label{change1}\, .
\eeq
The new coordinates in the configuration space are now
commutative still verifying $[Q_i,P_j]=i\delta_{ij}$, and the Hamiltonian gets the following form in terms of them:
\beq
H=\frac{\omega_+}{2}(P_1^2+Q_1^2)+ \frac{\omega_-}{2}(P_2^2+Q_2^2)\, ,
\eeq
where $\omega_\pm=\omega\lambda_\pm$. In this formulation (which we
will refer to as \emph{canonical formulation} in contrast to the path
integral) it is easy to see
therefore that the isotropic quantum oscillator of frequency $\omega$
on the noncommutative plane has the same spectrum as an anisotropic
oscillator on the commutative plane with frequencies
$\omega_{\pm}$~\cite{NCQMOsc,QTNCF}.

Defining the matrix $(\Omega^{-1})_{ij} = -i[\xi_i,\xi_j]$
is also useful to pass to the first order noncommutative Lagrangian
which is written as

\beq
\mathcal{L}_{nc}=\frac 12\Omega_{ij}\xi_i\dot \xi_j - H (\xi) =
\dot q_1 p_1 + \dot q_2 p_2 + \theta \dot p_2 p_1 - H(q_i,p_i)\, ,
\label{Omega}
\eeq
up to total derivatives.\footnote{In the former equation and from now
  on we will use the same notation for the phase space coordinates in
  the classical action as we have used for the quantum operators.}


\section{Noncommutativity in the path integral formulation}

We are going now to reformulate the problem of the quantum harmonic
oscillator on the noncommutative plane, which was discussed in the canonical
formulation in the previous section, by using the path integral
formalism. One has a sum over all paths in a four-dimensional
phase space, each one weighted by a phase factor (action) which depends on the
noncommutativity parameter~\cite{PINCQM}. This will allow us to
reinterpret the effect of the space noncommutativity as a mapping of paths.

The starting point is the expression for
the action of the two dimensional isotropic harmonic
oscillator in the commutative plane

\beq
S_c = \int d\tau {\cal L}_c(\tau) = \int_{-\infty}^{\infty}
\frac{d\epsilon}{2\pi\omega} {\cal L}_c(\epsilon) \,,
\label{Sc}
\eeq
with
\beq
{\cal L}_c(\epsilon) = -\frac{1}{2} \xi_c^\dagger(\epsilon) A_c(\epsilon)
\xi_c(\epsilon) \,,
\label{Lc}
\eeq
\beq
A_c(\epsilon) \,=\, I + i\frac{\epsilon}{\omega}\Omega(\theta=0) \,,
\eeq
$\xi_c$ is a matrix notation for the four phase space coordinates in
the commutative plane and
\beq
f(\epsilon) = \omega \int d\tau e^{i\tau\epsilon} f(\tau)\, ,
\eeq
with time parameter $\tau$.\footnote{We are going to assume that all
the integrals appearing throughout this work are sufficiently well
defined.}

The action in the noncommutative case can be expressed in a similar
way
\beq
S_{nc} = \int dt {\cal L}_{nc}(t) = \int_{-\infty}^{\infty}
\frac{dE}{2\pi\omega} {\cal L}_{nc}(E) \,,
\label{Snc}
\eeq
\beq
{\cal L}_{nc}(E) = -\frac{1}{2} \xi_{nc}^\dagger(E)
A_{nc}(E) \xi_{nc}(E) \,,
\label{Lnc}
\eeq
\beq
A_{nc}(E) \,=\, I + i\frac{E}{\omega}\Omega(\theta) \,.
\eeq
The subscript in $\xi_{nc}$ only emphasizes that it corresponds to the
column vector of phase space coodinates $\{\xi_i\}$ of the
noncommutative plane as defined in (\ref{change1}). For completeness
we give the explicit form of the matrix
$\Omega(\theta)$
\beq
\Omega(\theta) =  \left(\begin{array}{cccc} 0 & 0 & -1 & 0 \\
0 & 0 & 0 & -1 \\ 1 & 0 & 0 & \theta \\ 0 & 1 & -\theta & 0
\end{array}\right) \,.
\eeq
The notation for the time parameter is now $t$ and
\beq
f(E) = \omega \int dt e^{itE} f(t) \,.
\eeq

All the effect of the noncommutativity at this level is concentrated
in the phase space matrix $A_{nc}$ to be compared with $A_c$ in the
commutative case. The matrix $A_c$ has two (doubly) degenerate
eigenvalues $1+\epsilon/\omega$, $1-\epsilon/\omega$, and the
corresponding normalized eigenvectors are
\beq
v_1 = \frac 1{\sqrt{2}} \left(\begin{array}{c}  q_- \\ i q_-
\end{array}\right)   {\hskip 0,5cm}
v_2 = \frac 1{\sqrt{2}} \left(\begin{array}{c}  q_+ \\ i q_+
\end{array}\right)   {\hskip 0,5cm}
v_3 = \frac 1{\sqrt{2}} \left(\begin{array}{c}  q_- \\ -i q_-
\end{array}\right)   {\hskip 0,5cm}
v_4 = \frac 1{\sqrt{2}} \left(\begin{array}{c}  q_+ \\ -i q_+
\end{array}\right) \,,
\eeq
where $q_-$, $q_+$ are normalized two component column vectors
satisfying
\beq
\sigma_{2} q_{-} = - q_{-}  {\hskip 2cm}  \sigma_{2} q_{+} =  q_{+} \,,
\eeq
where
\beq
\sigma_{2} = \left(\begin{array}{cc}  0 & -i \\ i & 0 \end{array}\right)
\eeq
is the second Pauli matrix.

In the noncommutative case one has the eigenvalues and normalized eigenvectors of
the matrix $A_{nc}$
\beq
\left(1 + \lambda_+ \frac{E}{\omega}\right) {\hskip 2cm}
u_1 = \frac 1{\sqrt{1 + \lambda_+^2}}\left(\begin{array}{c}  q_- \\ i
\lambda_+ q_- \end{array}\right)
\eeq
\beq
\left(1 + \lambda_- \frac{E}{\omega}\right) {\hskip 2cm}
u_2 = \frac 1{\sqrt{1 + \lambda_-^2}} \left(\begin{array}{c}  q_+ \\ i \lambda_- q_+ \end{array}\right)
\eeq
\beq
\left(1 - \lambda_- \frac{E}{\omega}\right) {\hskip 2cm}
u_3 = \frac 1{\sqrt{1 + \lambda_-^2}} \left(\begin{array}{c}  q_- \\ - i \lambda_- q_- \end{array}\right)
\eeq
\beq
\left(1 - \lambda_+ \frac{E}{\omega}\right) {\hskip 2cm}
u_4 = \frac 1{\sqrt{1 + \lambda_+^2}} \left(\begin{array}{c}  q_+ \\ - i \lambda_+ q_+ \end{array}\right)\, .
\eeq

According to the signs of the eigenvalues of the matrix
$A_c(\epsilon)$ one has a decomposition of the commutative action
\beq
S_c = \int_{-\infty}^{-\omega} \frac{d\epsilon}{2\pi\omega} {\cal
L}_c(\epsilon) + \int_{-\omega}^{\omega} \frac{d\epsilon}{2\pi\omega} {\cal
L}_c(\epsilon) + \int_{\omega}^{\infty} \frac{d\epsilon}{2\pi\omega} {\cal
L}_c(\epsilon) \,.
\eeq
In the first term one has signs $(-,+)$ for the two doubly degenerate
eigenvalues of $A_c(\epsilon)$, in the second term both eigenvalues
are positive, and in the third term one has signs $(+,-)$.

A similar decomposition for the noncommutative action leads us to
\beq
S_{nc} = S_{nc}^{\theta} + {\bar S}_{nc}^{\theta}
\label{Stheta}
\eeq
with
\beq
S_{nc}^{\theta} = \int_{-\infty}^{-\omega/\lambda_-} \frac{dE}{2\pi\omega} {\cal
L}_{nc}(E) + \int_{-\omega/\lambda_+}^{\omega/\lambda_+} \frac{dE}{2\pi\omega} {\cal
L}_{nc}(E) + \int_{\omega/\lambda_-}^{\infty} \frac{dE}{2\pi\omega} {\cal
L}_{nc}(E)
\eeq
\beq
{\bar S}_{nc}^{\theta} = \int_{-\omega/\lambda_-}^{-\omega/\lambda_+}
\frac{dE}{2\pi\omega} {\cal L}_{nc}(E) + \int_{\omega/\lambda_+}^{\omega/\lambda_-}
\frac{dE}{2\pi\omega} {\cal L}_{nc}(E) \,.
\eeq
The three terms in $S_{nc}^{\theta}$ correspond to the combinations of
signs $(-,-,+,+)$, $(+,+,+,+)$ and $(+,+,-,-)$ respectively for the four
eigenvalues of $A_{nc}(E)$, and the two terms in ${\bar
S}_{nc}^{\theta}$ to $(-,-,-,+)$ and $(+,+,+,-)$.

The three terms in the commutative action can be mapped to the three terms in
$S_{nc}^{\theta}$ with the identifications $E=\epsilon/\lambda_-$,
$E=\epsilon/\lambda_+$ and $E=\epsilon/\lambda_-$. To make explicit
the mapping we introduce new phase space coordinates $a_i$ by
\beq
\xi_{nc}(E) = \sum_{i=1}^{4} a_i(E) u_i
\label{u}
\eeq
where $u_i$ are the four eigenvectors of $A_{nc}(E)$. In a similar way
one can use the four eigenvectors $v_i$ of the matrix $A_c(\epsilon)$
to introduce in the commutative case phase space coordinates $b_i$
through the expansion
\beq
\xi_c(\epsilon) = \sum_{i=1}^{4} b_i(\epsilon) v_i \,.
\label{v}
\eeq
Now we can define a mapping between paths in the phase spaces of the commutative and
noncommutative systems
\beq
\begin{array} {cc} a_1(\epsilon/\lambda_-) = \sqrt{\dfrac{\omega +
\epsilon}{\lambda_- \omega + \lambda_+ \epsilon}}\;\; b_1(\epsilon) &
a_2(\epsilon/\lambda_-) = \sqrt{\lambda_+} \;\; b_2(\epsilon) \\
a_3(\epsilon/\lambda_-) = \sqrt{\lambda_+} \;\; b_3(\epsilon) &
a_4(\epsilon/\lambda_-) = \sqrt{\dfrac{\omega -
\epsilon}{\lambda_- \omega - \lambda_+ \epsilon}}\;\; b_4(\epsilon) \end{array}
\label{b2a>}
\eeq
when $|\epsilon| > \omega$ and
\beq
\begin{array} {cc}
a_1(\epsilon/\lambda_+) = \sqrt{\lambda_-} \;\; b_1(\epsilon) &
a_2(\epsilon/\lambda_+) = \sqrt{\dfrac{\omega +
\epsilon}{\lambda_+ \omega + \lambda_- \epsilon}}\;\; b_2(\epsilon) \\
a_3(\epsilon/\lambda_+) = \sqrt{\dfrac{\omega -
\epsilon}{\lambda_+ \omega - \lambda_- \epsilon}}\;\; b_3(\epsilon) &
a_4(\epsilon/\lambda_+) = \sqrt{\lambda_-} \;\; b_4(\epsilon) \end{array}
\label{b2a<}
\eeq
when $|\epsilon| < \omega$.
With this mapping one has
\beq
S_{nc}^\theta[a_i] = S_c[b_i] \,.
\eeq
The decomposition of the noncommutative action as a sum of two
contributions (\ref{Stheta}) can be translated at the level of
phase space paths
\beq
\xi_{nc}(t) = \xi_{nc}^{\theta}(t) + {\bar \xi}_{nc}^{\theta}(t)
\label{decomp}
\eeq
with
\beq
\xi_{nc}^{\theta}(t) = \int_{-\infty}^{-\omega/\lambda_-}
\frac{dE}{2\pi\omega} e^{-iEt} \xi_{nc}(E)
+ \int_{-\omega/\lambda_+}^{\omega/\lambda_+} \frac{dE}{2\pi\omega}
e^{-iEt} \xi_{nc}(E)
+ \int_{\omega/\lambda_-}^{\infty} \frac{dE}{2\pi\omega}
e^{-iEt} \xi_{nc}(E)\, ,
\label{xinctheta}
\eeq
\beq
{\bar \xi}_{nc}^{\theta}(t) =
\int_{-\omega/\lambda_-}^{-\omega/\lambda_+}
\frac{dE}{2\pi\omega} e^{-iEt} \xi_{nc}(E)
 + \int_{\omega/\lambda_+}^{\omega/\lambda_-}
\frac{dE}{2\pi\omega} e^{-iEt} \xi_{nc}(E) \,.
\eeq
Since every path in phase space $\xi_{nc}$ is decomposed into a
a direct sum of two paths $\xi_{nc}^{\theta}$ and $ {\bar
  \xi}_{nc}^{\theta}$, this amounts to a decomposition of the space of paths.

The mapping between paths can then be re-expressed in the form
\beq
\xi_{nc}^{\theta}(t) = \omega \int d\tau \;\; K_{\xi}(t,\tau) \;\;
\xi_c(\tau) \,.
\label{map}
\eeq
The explicit form of the kernel matrix $K_{\xi}(t,\tau)$ can be obtained by
combining (\ref{xinctheta}) with the expansions of $\xi_{nc}(E)$ and
$\xi_c(\epsilon)$ in eigenvectors of the matrices $A_{nc}(E)$ and
$A_c(\epsilon)$ (\ref{u}-\ref{v}) and the mapping from $a_i$ to $b_i$
(\ref{b2a>}-\ref{b2a<}).

To summarize one has a noncommutative action
\beq
S_{nc}[\xi_{nc}] =  S_{nc}^{\theta}[\xi_{nc}^{\theta}] + {\bar
S}_{nc}^{\theta}[{\bar \xi}_{nc}^{\theta}] =  S_{c}[\xi_{c}] + {\bar
S}_{nc}^{\theta}[{\bar \xi}_{nc}^{\theta}] \,.
\eeq
The effect of the noncommutativity in the path integral formulation is
twofold: a mapping of paths (\ref{map}) and the addition of a term in
the action (${\bar S}_{nc}^{\theta}$).

A similar analysis of the effect of noncommutativity can be made at
the level of paths in configuration space. One can repeat step by step
the discussion of the path integral formulation in phase space. We
give directly the result. One has
\beq
S_{nc}[q_{nc}] = S_{nc}^{\theta}[q_{nc}^{\theta}] + {\bar
S}_{nc}^{\theta}[{\bar q}_{nc}^{\theta}]
\eeq
with
\beq
q_{nc}(t) = q_{nc}^{\theta}(t) + {\bar q}_{nc}^{\theta}(t)
\eeq
\beq
q_{nc}^{\theta}(t) = \int_{- \omega/\theta}^{\infty}
\frac{dE}{2\pi\omega} e^{-iEt} q_{nc}(E) {\hskip 1cm}
{\bar q}_{nc}^{\theta}(t) = \int_{-\infty}^{- \omega/\theta}
\frac{dE}{2\pi\omega} e^{-iEt} q_{nc}(E) \,.
\eeq
The map
\beq
q_{nc}(E) = \sqrt{\frac{dE}{d\epsilon}} \;\; q_c(\epsilon)
\eeq
with
\beq
\cfrac{E}{\sqrt{1+\theta\cfrac{E}{\omega}}} = \epsilon
\eeq
for $E>-\omega/\theta$ leads to
\beq
S_{nc}^{\theta}[q_{nc}^{\theta}] = S_c[q_c]
\eeq
where $S_c$ is the action in the commutative case that corresponds to
the path
\beq
q_c(\tau) = \int_{-\infty}^{\infty} \frac{d\epsilon}{2\pi\omega}
e^{-i\epsilon\tau} q_c(\epsilon) \,.
\eeq
Then once more one has that the effect of the noncommutativity at the
level of the path integral formulation in configuration space is a
mapping of paths
\beq
q_{nc}^{\theta}(t) = \omega \int d\tau \;\; K_q(t,\tau) \;\; q_c(\tau)
\eeq
and the addition of a term in the action (${\bar S}_{nc}^{\theta}$).

The characterization of noncommutativity as a decomposition of the
space of paths~(\ref{decomp}), together with a mapping of the space of paths $\xi_c$ into the
subspace of paths $\xi_{nc}^{\theta}$ corresponding to one of the components of the
decomposition, is a consequence of the path integral formulation of the
quantum system. In the canonical formulation of the previous section
one had just a (noncanonical) change of variables in phase space
associated to the noncommutativity of space. In fact this change of
variables could have been introduced also at the level of the action
as a functional defined in the space of paths to show that it
can be written as a sum of two actions, each one corresponding to a one
dimensional harmonic oscillator of different frequencies.


\section{Summary}

We have shown that for a quantum linear system (quadratic action) it
is possible to identify a mapping of paths that allows to go partially
from the quantum system defined in a commutative space to the system
in a noncommutative space. We say partially because the noncommutative
action can be written as a sum of two independent contributions (in
the sense that they involve different decoupled degrees of freedom)
and it is only one of them that can be obtained from the commutative
action through a mapping of paths.

The identification of this mapping associated to noncommutativity
opens a new way to introduce nonlinear effects. Instead of including
directly non-quadratic terms at the level of the noncommutative action,
one can apply the mapping identified in the linear system to a
commutative action including non-quadratic terms, and then add the quadratic 
additional action $\bar S^{\theta}$ of the linear system to the resulting action. In
this way one generalizes the correspondence of the commutative and
noncommutative actions to nonlinear systems. It seems interesting to
investigate the consistency and properties of a quantum system
defined in this way.

The possibility to use an alternative characterization of
space noncommutativity to introduce nonlinear effects in a different
way while keeping a simple relation with the commutative case can be
just a curiosity at the level of quantum mechanics but the extension
of the discussion presented in this work to quantum
field theory (\cite{futuro}) may be essential to formulate a theory
with interactions which is consistent in the presence of
noncommutativity at the level of fields.


\ack This work has been partially supported by CICYT (grant
FPA2006-02315) and DGIID-DGA (grant2008-E24/2). J.I. acknowledges a FPU
grant and D.M. a FPI grant from MICINN.

\end{document}